\documentclass[twocolumn]{aastex63}
\usepackage{amsmath}
\usepackage{natbib}
\usepackage[version=4]{mhchem}
\usepackage{color}
\newcommand{\md}{\text d}
\newcommand{\species}{\ce{H2O}, \ce{CH3OH}, and \ce{H2CO} }
\def\xian#1{#1}

\shorttitle{An active Sgr A* and the synthesis of water and organic molecules in the MW}
\shortauthors{Chang Liu, Xian Chen, \& Fujun Du}

\begin{document}
\title{Impact of an Active Sgr A* on the Synthesis of Water and Organic Molecules Throughout the Milky Way}
\author{Chang Liu}
\affiliation{Department of Astronomy, School of Physics, Peking University, Beijing 100871, China}
\author{Xian Chen}
\email{Corresponding author: xian.chen@pku.edu.cn}
\affiliation{Department of Astronomy, School of Physics, Peking University, Beijing 100871, China}
\affiliation{Kavli Institute for Astronomy and Astrophysics, Beijing 100871, China}
\author{Fujun Du}
\affiliation{Purple Mountain Observatory, Chinese Academy of Sciences, Nanjing 210008, China}

\begin{abstract}
Sgr A*, the supermassive black hole (SMBH) in our Galaxy, is dormant today, but
it should have gone through multiple gas-accretion episodes in the past 
billions of years
to grow to its current mass of $4\times10^6\,M_\odot$.  Each episode
temporarily ignites the SMBH and turns the Galactic Center into an
active galactic nucleus (AGN). Recently, we showed that the AGN could
produce large amount of hard X-rays that can penetrate the dense
interstellar medium in the Galactic plane. Here we further study the
impact of the X-rays on the molecular chemistry in our Galaxy. We use a
chemical reaction network to simulate the evolution of several
molecular species including \ce{H2O}, \ce{CH3OH}, and \ce{H2CO},
both in the gas phase and on the surface of dust grains.  We find that
the X-ray irradiation could significantly enhance the abundances of
these species. The effect is the most significant in those young,
high-density molecular clouds, and could be prominent at a Galactic distance of
$8$ kpc or smaller.  The imprint in the chemical abundance is visible even
several million years after the AGN turns off. 
\end{abstract}

	
\section{Introduction}
\label{sec:Intro}

Supermassive black holes (SMBHs) are ubiquitous in the centers of massive
galaxies \citep{Kormendy2013}.  In theory, a SMBH grows to its current mass
mainly through multiple episodes of gas accretion \citep{volonteri10}.  In each
accretion phase a significant fraction of the gravitational energy of the gas
is released in the form of radiation, and the galaxy center consequently
becomes an active galactic nucleus \citep[AGN,][]{soltan82}.  Analysis of the
luminosity function of bright AGNs indicates that the accretion episodes may
add up to about $(1-10)\%$ of the lifetime of a galaxy
\citep{hopkins09,shankar09}. The SMBHs in the mass range of
$(10^6-10^7)\,M_\odot$ on average are more active in the sense that they become
AGNs every $10^7-10^8$ years while each active phase only lasts $10^5-10^6$
years \citep{hopkins06}.

The center of the Milky Way (MW) also harbors a SMBH. It coincides with the
bright radio source Sagittarius A* (Sgr A*) and the mass is estimated to be
$4\times10^6\,M_\odot$ \citep{Genzel2010}.  This black hole (BH) is dormant
today, but in the past it should have been active according to the
close relationship between SMBH growth and AGN activity \citep{mezger96}.  
One
interesting discovery in the last decade is that Sgr A* may be an AGN just
several million years (Myrs) ago. This picture is supported by a series of
findings, including the Fermi bubble \citep{su10}, the bright emission of
H$\alpha$ lines in the Magellanic Stream \citep{BlandHawthorn13}, and the young
stellar disk around Sgr A* which is reminiscent of a relic accretion disk
\citep[see][for a review]{ChenAmaro-Seoane2014b}.  The peak luminosity of that
recent AGN is unclear and the current estimations fall in a broad range of
$(3-100)\%$ of the Eddington limit $L_{\rm Edd}\simeq5\times10^{44}\,{\rm
erg\,s^{-1}}$ \citep{nay05,BlandHawthorn13}.  Such a high level of activity is
not completely unexpected given the aforementioned frequent activity of
low-mass BHs. 

Not many works have explored the impact of an active Sgr A* on the habitability
of the MW.  Early studies identified X-ray and cosmic rays as a potential
threat to lives but they lacked a theoretical framework to quantify the damage
\citep{clark81,laviolette87,gonzalez05}.  \citet[][Paper I]{amaro-seoane19}
adopted an empirical AGN spectral energy distribution (SED) and calculated the
extinction of light in the Galactic plane. They found that hard X-rays ($>2$
keV) could reach Earth unattenuated and the corresponding flux is comparable to
an X-class solar flare.  They also suspected that the high X-ray irradiation
may have a noticeable impact on the molecular chemistry in the MW.  Other works
neglected the attenuation of light and focused on the evaporation of planet
atmosphere during the AGN irradiation
\citep{balbi17,chen18habitability,forbes18,wislocka19}.  Their general
conclusion is that mass loss is significant only within a distance of $1$ kpc
from Sgr A*. More recently, \citet{lingam19} explored possible beneficial
effects associated with the AGN irradiation, such as a prebiotic synthesis of
the building blocks of biomolecule and a powering of photosynthesis on
free-floating planets. Based on the estimation of the UV flux, they reached a
similar conclusion that only in the central kpc of the MW are these effects
important. 

In this work, we continue our early study of the impact of an active Sgr A* on
the habitability of the MW. We focus on hard X-ray irradiation because, unlike
optical and UV, hard X-ray photons are not attenuated by the gas in the
Galactic plane and hence could reach large distances (see Paper I).  It is
known, from studying the molecular clouds in star-forming regions, that hard
X-rays could enhance the abundance of organic molecules such as \ce{CN,\ C2H,\
HCN} by $2-6$ orders of magnitude \citep{Krolik1983,lepp96}. The main
reason is that X-rays, by ionizing atoms and molecules, create energetic
electrons \citep{Shull1985,Maloney1996}, which, by further interacting with the
atoms and molecules,  produce reactive ions and radicals \citep{Herbst1973,
Herbst2009}.  Ions and radicals, compared to neutral species, react more
quickly to produce complex molecules \citep{Herbst2009,Wakelam2010}. 

Because of the close relationship between X-ray irradiation and the synthesis
of organic molecules, we expect an imprint of the recent active Sgr A*  in the
abundance of complex molecules in the MW.  This is a reasonable expectation
also because abnormal molecular \xian{abundances have} been detected in external
galaxies with AGNs \citep{Usero2004}.  For example, the column densities of
\ce{H2O} and some organic species such as \ce{HCN}, \ce{HCO+}, and
\ce{CH3OH} are particularly high in AGNs
\citep{Gonzalez-Alfonso2010,Davies2011,Imanishi2014,takano14,Harada2018}
\xian{compared to those in MW clouds (see below)}.
The canonical explanation of this enhancement is that the AGN produces an
X-ray dominated region (XDR) within a distance of $10-10^2$ pc
\citep{Meijerink05,meijerink07,viti14,izumi15}.  At such a close distance, the
X-ray flux could be as high as $10^2\,{\rm erg\,cm^{-2}\,s^{-1}}$, which
exceeds the UV radiation produced by the starburst in a galactic nucleus
\citep[typically $10\,{\rm erg\,cm^{-2}\,s^{-1}}$,
e.g.,][]{Garcia-Burillo10,izumi15}.  As a result, the molecular clouds in this
region are ionized mainly by X-rays and heated to a temperature of $100-200$ K.
Under these conditions, the synthesis of \ce{H2O} and organic molecules can be
enhanced. \xian{Besides X-ray irradiation, mechanical heating due wind or jet also plays an
important role in the enhancement of organic molecules, such as \ce{HCN} \citep{Garcia-Burillo14,izumi15,martin15}. }

Although these previous studies revealed a positive correlation between
X-ray irradiation and the abundances of water and organic molecules, we cannot
draw directly the conclusion that the past activities of Sgr A* have had a
positive impact on the production of water and organic molecules in the MW. The
reasons are as follows. (i) We are interested in the molecular clouds
throughout the MW, with the distance from the Galactic Center ranging from
hundreds of parsecs to ten kpc. In this distance range, the X-ray flux
impinging on the clouds is not as strong as those shown above in AGNs. For
example, in Paper I we showed that at the location of the solar system, which
is $8$ kpc from the Galactic Center, the X-ray flux during the active phase of
Sgr A* is a few times $10^{-3}\,{\rm erg\,cm^{-2}\,s^{-1}}$.  Such a flux is
not much higher than the UV-background radiation in the Galactic plane
\cite{Draine1978}.  Therefore, our molecular clouds are not in the XDR. In
fact, neither the UV background nor the cosmic rays can be neglected in the
chemistry model of these clouds.  (ii) When the X-ray flux is as low as
$0.1\,{\rm erg\,cm^{-2}\,s^{-1}}$, the molecular clouds would not be
significantly heated by the X-ray ionization and could remain cold ($<50$ K)
\citep{meijerink07}. In such a ``cold cloud'', the dominant chemical reactions
and the reaction rates differ from those in an XDR \citep{Herbst2009}. It is
worth noting that cold clouds represent the initial conditions for the
formation of low-mass stars, such as our Sun, and may determine the early
chemical composition of solar-like systems.  Therefore, it is important to
understand the potential impact of the X-ray irradiation on the synthesis of
water and organic molecules in them.

In cold clouds, several molecules are synthesized at a relatively early
stage, some of which are the building blocks of more complex, larger molecules.
Among these ``zeroth-generation'' species, water (\ce{H2O}), methanol
(\ce{CH3OH}), and formaldehyde (\ce{H2CO}) are commonly found in the molecular
clouds of the MW \citep{Herbst2009}. Therefore, they are particularly
interesting to our study. (i) Water is an important molecule involved in the
processes leading to the origin of life.  It can be synthesized in gas as well
as on the surface of dust grains. In cold clouds,  the water vapor abundance
(relative to \ce{H2}) is normally $10^{-10}-10^{-8}$ \citep{snell00}, while the
surface abundance of water ice is much higher, about $10^{-4}$ of \ce{H2}
\citep{Bergin07}.  Around AGNs, the abundance of water vapor can be as high as
$10^{-6}$ \citep{Gonzalez-Alfonso2010}, confirming the positive effect of X-ray
irradiation on the synthesis of water molecules.  (ii) Methanol (\ce{CH3OH}) is
one of the most common organic molecules detected in cold  clouds
\citep{herbst09}. Its abundance varies from $10^{-9}$ and below in the gas
phase ($10$ K) to as high as $10^{-6}-10^{-5}$ in ice \citep{herbst09}.  In
AGNs, the gas abundance is typically $10^{-7}$
\citep{Garcia-Burillo10,nakajima15,Harada2018}. These \ce{CH3OH} in XDRs may be
produced on grains and then released into the gas due to shock.  (iii)
Formaldehyde (\ce{H2CO}) is often detected at the surface of a cold cloud.
This species  may be produced by grain processes deep in the cores of the cloud
and later brought to the cloud surface by turbulence \citep{federman91}. In
cold clouds,  the typical gas abundance is $10^{-10}-10^{-9}$ \citep{carey98},
but in XDRs the abundance can be $1-2$ orders of magnitude higher
\citep{Harada2015,Harada2018}.

The chemical process which lead to the formation of large, organic
molecules in the MW is still under intensive investigation, and the role of the
X-rays from Sgr A* is unclear.  As a first step towards understanding their
possible relationship, we study in this work the synthesis of water and the
``zeroth-generation'' organic molecules in cold clouds. We pay particular
attention to \ce{CH3OH} and \ce{H2CO}, for the reasons given above, but we also
study several other organic molecules, such as \ce{HCN}, \ce{HCO+}, which are
known to be enhanced in XDRs
\citep{kohno03,Usero2004,Davies2011,Imanishi2014,viti14,martin15}.

This paper is organized as follows. In \S\ref{sec:Methods} we calculate the
X-ray irradiation spectra at different distances from Sgr A*, and we also
specify our X-ray chemical model, especially the processes associated with
\species.  In \S\ref{sec:Simulations} we describe our simulation of the
molecular chemistry induced by X-ray irradiation.  The results are presented in
\S\ref{sec:Results}, where we show the dependence of the abundance of different
molecules on the distance, column density, and lifetime of a molecular cloud.
A possible diagnostics is provided in \S\ref{sec:diag} for future observational test
of our theoretical results.
We discuss the caveats of our current model in \S\ref{sec:caveat} and finally
summarize our conclusions in \S\ref{sec:Summary}.

\section{X-ray Chemistry in Molecular Clouds} \label{sec:Methods}

\subsection{X-ray Source} \label{sec:X-ray} 

\begin{figure} 
\centering
\includegraphics[width=\linewidth]{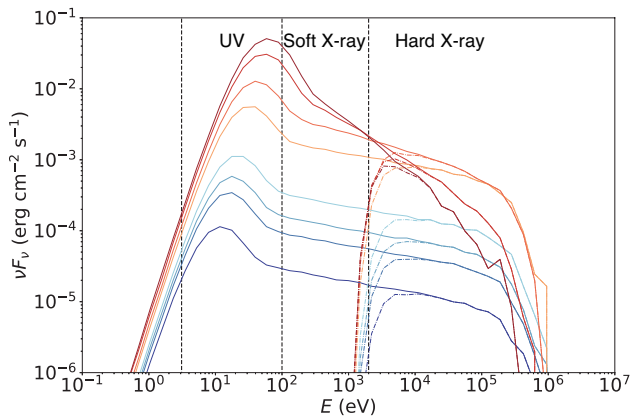} \caption{ The SED of Sgr A*
	at a Galactic distance of $8$ kpc.  The solid curves show the
	unabsorbed SEDs of the AGN assuming a bolometric luminosity of $(3.0,
	2.0, 1.0, 0.5, 0.1, 0.05, 0.03, 0.01)$ times the Eddington luminosity. The
dot-dashed ones account for the extinction at a Galactic distance of $8$ kpc.
} \label{fig:1} \end{figure}

To calculate the X-ray flux during the AGN phase of Sgr A*, we use a numerical
model which is described in detail in \cite{Liu2016}.  This model computes the
SED of an accretion disk based on three parameters, the Eddington ratio defined
as $\eta:=L/L_{\rm Edd}$ where $L$ is the bolometric luminosity of the disk,
the X-ray spectral index $\alpha$, and the magnetic parameter $\beta$ which
characterizes the relative importance between the sum of the gas and radiative
pressure and the magnetic pressure. In this model, hard X-rays are produced by
a hot corona screening the disk. The solid curves in Figure~\ref{fig:1} show
the output SED as a function of the Eddington ratio. In the calculation we
assume that $\alpha=0.1,\ \beta=100$, and the BH mass is
$4\times10^6\,M_\odot$.  We find that when $\eta\ga1$, the X-ray flux
above $2$ keV is more or less constant, at a level of a few times $10^{-3}~{\rm
erg\,s^{-1}\,cm^{-2}}$. This result from numerical simulation is consistent
with our analytical estimation in Paper I.  When $\eta\la0.5$, we find that the
luminosity of hard X-ray falls linearly with decreasing Eddington ratio.

Soft photons are subject to extinction as they propagate in the Galactic plane.
To quantify it, we adopt the empirical density profile of the atomic
hydrogen $\rho_{\ce{H}}(R)$ and molecular hydrogen $\rho_{\ce{H2}}(R)$ in the
MW \citep{McMillan2017}, and we use the software
\texttt{Xspec}\footnote{\url{https://heasarc.gsfc.nasa.gov/xanadu/xspec/}} to
calculate the extinction. Since most molecular clouds are inside the Galactic
plane, we only consider the horizontal (in-plane) gas distribution. The corresponding
column density at a Galactic distance of $R$ is
\begin{equation} 
\Sigma(R)=\int_{0}^{R}\left[\rho_{\ce{H}}(x)+\rho_{\ce{H2}}(x)\right] \md x. 
\end{equation}
The residual SEDs at $R=8$ kpc, for example, are shown in  Figure~\ref{fig:1}
as the dot-dashed curves.  We can see that the optical and UV radiation is
completely absorbed. For this reason, in our chemistry model we focus on
the effects induced by X-rays.

\subsection{X-ray Ionization}\label{sec:ionization}

It is mainly through ionization that X-ray irradiation could affect the chemistry inside a molecular cloud \citep{Maloney1996}. Other effects, such as Coulomb heating, are less
important in an environment where the ionization fraction $x_e$ is relatively
low. In our model we consider two types of ionization.

Primary ionization, also known as the direct
photoionization, is caused by an X-ray photon striking an atom
\citep[e.g.][]{Latif2015}.
The primary ionization rate for a given species  
$i$ can be calculated with 
\begin{equation} \label{eq:1}
\zeta_p^i=\int_{E_{min}}^{E_{max}}\frac{F(E)}{E}e^{-\tau(E)}\sigma^i(E)\text{d}E, 
\end{equation} 
where $F(E)$ is the monochromatic X-ray flux incident on the surface of a
molecular cloud, $\sigma^i(E)$ is the photoionization cross section given a
photon energy of $E$ \citep{Verner1996}, $\tau$ is the optical depth
defined as
\begin{equation} \label{eq:optical_depth} \tau(E)=\sum_{i=\ce
H,\ce{He}}N_i\sigma^i(E), 
\end{equation}
and $N_i$ is the column density of a species.  In calculating the optical
depth, we have taken into account the fact that \ce{H} (mostly in
molecular hydrogen) and \ce{He} contribute to most of the opacity.

The photoelectrons released in primary ionization could further collide with other
atoms and cause secondary ionization \citep{Maloney1996,
Stauber2005}. In principle, the secondary ionization rate can be calculated with
\begin{equation}
\zeta^i_{\rm sec}=\int_{E_{min}}^{E_{max}}\frac{F(E)}{E}e^{-\tau(E)}N_{\rm sec}(E,x_e)\sigma^i(E)\text{d}E, 
\end{equation}
where $N_{\rm sec}(E,x_e)$ is the number of secondary ionization events produced by
each photoelectron in average. If the energy of the photoelectron is $E$, this number can be calculated with
\begin{equation}
	N_{\rm sec}(E,x_e)=\frac{\eta_{\rm ion}(x_e) E-E_{\rm th}}{W(E)}, 
\end{equation}
where $\eta_{\rm ion}$ is the fraction of the
photoelectron's energy which goes into secondary ionization, $E_{\rm th}$
is the ionization threshold of an atom, and $W$ is the mean energy needed to
produce an ion-electron pair.  In practice, we simplify the calculation by
adopting the ionization rates presented in \cite{Shull1985}.  These rates are
derived from a Monte Carlo simulation of the secondary ionization of \ce{H}
and \ce{He} atoms over a wide range of electron fraction ($0.0001<x_e<1$) and
photoelectron energy (from $100$ eV to several keV).

Although we have two types of ionization in our model, the secondary ionization
rate predominates. For example, if we consider a typical value of $x_e\le1\%$
for a cold ($T\sim10$ K) molecular cloud, approximately $40\%$ of the energy of
primary photoelectrons goes into secondary ionization.  Since the ground state
of neutral \ce{H} has a ionization threshold $E_{\rm th}=13.6\text{ eV}$,
using $\eta_{\rm ion}=40\%$ and $W(E)=E_{\rm th}$, we find that for a typical
primary photoelectron with $E=1$ keV, the number for secondary ionization
events it will cause is $N_{\rm sec}(E,x_e)\approx30$.

For \ce{H2}, \ce{H} and \ce{He}, we consider both primary and secondary
ionization.  For heavy elements (\ce{C,\ N,\ O}, etc.) and molecules (\ce{CO,\ CO2}, etc.), we
consider only the secondary ionization for simplicity.  We assume that the
ionizing electrons are mainly from the three most abundant neutral species, i.e.,
\ce{H2}, \ce{H} and \ce{He}. Then the ionization rate of heavy elements
and molecules can be derived from
\begin{equation} \zeta ^ { m } =\frac { \sigma _ {
\mathrm { ei } ,m} ( E ) } { \sigma _ { \mathrm { ei } , \mathrm { H } } ( E )
} \zeta ^ { \ce{H} }_{\rm sec}, 
\end{equation}
where $\sigma_{\text{ei}}(E)$ is the electron-impact cross section at energy
$E$ \citep{Maloney1996,Adamkovics2011}.  We note that heavy elements are
usually bound in molecules, but the current derivation of
$\sigma_{\text{ei}}(E)$ assumes that the cross section  is unaffected by the
molecular binding energy.  Moreover, we notice that the ratio of the cross
sections, which appears in the last equation, is insensitive to $E$
\citep{Maloney1996}. For this reason, we use the cross sections at $E=50\text{
eV}$, which is the typical energy of the secondary electrons.

For completeness, we also consider the destruction of
\ce{H2} and several other molecules
caused by
photoionization. 
For example, photoionization of \ce{H2} can have two possible outcomes, pure
ionization and dissociation, which can be expressed as
\begin{align*} 
	\ce{H2 &-> H2+ + e-},\\ 
	\ce{H2 &->H+ + H + e-}. 
\end{align*} 
We take the branching ratio suggested by \cite{Krolik1983}, that the fraction
of pure ionization is approximately $80\%$ and that for dissociation is $
20\%$. For other atomic and molecular reactants, we adopt the same products
from the OSU chemical database
osu\_01\_2007\footnote{\url{https://faculty.virginia.edu/.archived/ericherb/research.html}},
as well as the corresponding branching ratios for cosmic-ray ionization and
dissociation. We choose this database because it was used by
\citet{Wakelam2008} to explained the molecular abundances observed in the cold
dense cores in the MW. These cores represent the intial conditions of the
moeluclar clouds of our own interest.  To be able to reproduce the results in
\citet{Wakelam2008}, we did not choose the latest chemical database provided by
the OSU group. The later updates included more anions, which are more important
for the formation of long carbon cahins \citep{Herbst2009}.

\subsection{Grain Processes}\label{sec:grain} 

The surface of a dust particle (grain) is an important place for the synthesis
of complex molecules. A grain accretes from the surrounding gas many molecular
species to its surface, where these molecules can interact more frequently and
the chemical reactions such as hydrogenation can proceed more efficiently.
Since we are interested in the synthesis of \ce{H2O}, \ce{CH3OH}, and \ce{H2CO}
(see \S\ref{sec:Intro}), we include in our model the surface reactions
containing the species made of \ce{H,\ C,\ N,\ O}. Only those species
with no more than two \ce{C} atoms are selected since we do not study
long-carbon-chain species in this work. Table~\ref{table:Grain_Reaction} in
\ref{sec:surface_reaction} shows the relevant reactions, which are selected from the
network presented in \cite{Hasegawa1992}. The last column shows the
activation energies for those reactions with high energy barriers. This
network is sufficient to simulate the species we are interested in.  The
reaction parameters in this network are still frequently used today, although
there are works which extend the reaction mechanisms so that the desorption
energy may depend on the fraction of surface coverage by molecules
\citep{garrod11} and that the molecules in ice mantle could react with each
other \citep{chang14}.  More recent grain-surface chemical networks
\citep[e.g.,][]{garrod08} are usually concerned with the formation of more
complex organic molecules, which is not the focus of the current work.  On one
hand, the abundances of those very complex molecules are usually low, so that
including them in our network should have a negligible effect on the species
studied in this paper. On the other, it may be interesting to ask to what
extent those more complex molecules are affected by the mechanisms discussed in
the current paper, which may be a topic for future investigation.

We calculate the accretion rate of molecules onto dust grains following the
method developed in \cite{Hasegawa1992}.  We consider accretion, via weak van
der Waals forces (physisorption), onto ``classic'' dust grains, which have a
fixed radius $r_d=1000\ \overset{\circ}{\text{A}}$, a density of $\rho=3$
g/cm$^3$, and $N_s=10^6$ sites for adsorption (the number of molecules that can
be absorbed to and held on a dust grain).  The dust temperature $T_d$ is
assumed to be the same as the gas temperature $T\sim10$ K, following the
assumption in \citet{Hasegawa1992}.  This is a reasonable approximation for the
clouds at $\ga2$ kpc but may be too crude for those at $1$ kpc. This is because
within $1$ kpc from the AGN, the X-ray flux significantly exceeds $0.1\,{\rm
erg\,s^{-1}\,cm^{-1}}$ so that heating due to ionization cannot be neglected
given our cloud density of $10^4\text{cm}^{-3}$ \citep{meijerink07}. Moreover,
we did not consider the difference between the gas and dust temperatures
because the difference is small due to the efficient gas-dust coupling at a
high density of $10^4\text{cm}^{-3}$ \citep{goldsmith01}.  Moreover, the
reaction rates are insensitive to small temperature changes (at least in the
range of temperature of our interest). For the gas-to-dust ratio in mass, we
adopt the standard value of $100$. We assume that the velocities of the gas
species obey Maxwell distribution. For any species that strikes a dust grain, a
sticking probability of $1$ is assumed.

For the desorption of molecules, we consider three channels, namely, thermal
desorption, cosmic-ray desorption, and photo-desorption. (i) Given the
adsorption energy $E_D$ and dust temperature $T_d$, the thermal evaporation
rate can be calculated with $R_{\text{evap}}\propto E_D^{1/2}\exp[-E_D/kT_d]$
\citep{Hasegawa1992}, where $k$ is the Boltzmann constant. The value of $E_D$
for each species is from \cite{Allen1977} and \cite{Hasegawa1993}. We note that
although thermal evaporation is common, the rate is negligible for any species
heavier than \ce{He} because of the low temperature in our molecular clouds.
(ii) The cosmic-ray induced desorption rates are calculated as in
\citep{Hasegawa1993}.  Every time a cosmic-ray particle strikes a dust grain,
the dust temperature is assumed to rise to $70$ K immediately and then drop via
thermal desorption.  (iii) Photo-desorption can be induced by either UV or
X-ray photons. For UV-induced desorption, we assume a rate of $10^{-3}$
molecule per grain per incident UV photon for any species in our grain-surface
network \citep[also see][]{Visser2011}.  Since UV-induced desorption depends
strongly on the visual extinction, which is a function of the optical depths of
a molecular cloud, we will study in \S\ref{sec:Av} the dependence of the rate
on the column density of the molecular cloud.  The X-ray desorption processes
are more complex (see \citealp{Jimenez-Escobar2018} for a brief review) and not
as well understood both in theory and in experiment. However, X-ray photons are
more penetrative than UV photons, so that the interaction happens deeper inside
the bulk of a grain and hence  does not as often lead to desorption. Therefore, the
X-ray desorption is less significant compared to UV photo-desorption and we
neglect it in our model.

\section{Simulation} \label{sec:Simulations}

We solve the evolution of the chemical network using the  public package
\texttt{KROME}\footnote{\url{http://kromepackage.org/}} \citep{Grassi2014}.
The initial conditions are as follows.  Our molecular cloud has a gas
temperature of $T=10\text{ K}$ and a density of
$n_{\ce{H}}=2\times10^4\text{cm}^{-3}$.  The initial abundances of elements are
taken from the EA2 model in \cite{Wakelam2008}, which is based on the
high-metal environment observed in the diffuse cloud of $\zeta$ Ophiuchi and is
modified based on recent observations of cold cores.  For the species at
different depths of a molecular cloud, we assume that they follow the same
initial abundances.

In our model, even when there is no X-ray irradiation, we include a low level
of ionization caused by cosmic rays. The corresponding ionization rate is
$\zeta=1.3\times10^{-7}$ s$^{-1}$ per hydrogen atom \citep{Wakelam2008}.  We
also include a generic UV background according to \cite{Draine1978}. Given the
UV flux, we calculate the photo-ionization and photo-desorption rates following
the scheme presented in \S\ref{sec:Methods}.

To simulate the impact of the AGN, 
we turn on X-ray chemistry in the network according to the method described in
\S\ref{sec:ionization} and \S\ref{sec:grain}. The X-ray SED is calculated
assuming a fiducial Eddington ratio of $\eta=1$. The extinction of optical, UV,
and soft-X-ray photons by the
Galactic plane is taken into account following \S\ref{sec:X-ray}.  
Since hard X-rays have relatively low extinction in the Galactic plane, 
the flux decreases 
with the distance approximately as $R^{-2}$. The corresponding values are
$(660,\,120,\,20,\,4.2)\times10^{-3}\,{\rm erg\,cm^{-2}\,s^{-1}}$ at $R=(1,\,2,\,4,\,8)\,{\rm kpc}$.
The
duration of the X-ray irradiation is set to $10^6$ years, to be consistent with
the empirical evidence (\S\ref{sec:Intro}).  After that, we turn off the X-ray
and let the network evolve for another $10^7$ years.  By the end of the
simulation, we investigate the abundance of the molecular species at different
Galactic distances to look for possible imprints of an AGN. In
our fiducial model we use $R=4$ kpc (\S\ref{sec:Default}). We also run
simulations using $R=(1,\,2,\,8)$ kpc for comparison (\S\ref{sec:Distribution}).

We notice that as the system evolves, the chemical abundance is no longer
uniformly distributed inside a molecular cloud because the extinction of UV and X-ray
radiation varies at different depth of the cloud. The depth $h$ can be characterized
by the column density of hydrogen, $N_{\ce{H}}$, which is related to $h$ as
$N_{\ce{H}}=n_{\ce{H}}\,h$. Therefore, to understand the abundance of molecular species
at different depth in a cloud, we run simulations using different $N_{\ce{H}}$.  
Given the value of $N_{\ce{H}}$, 
the extinction of X-ray inside the cloud is calculated using optical depth defined in
Equation~(\ref{eq:optical_depth}).
For the extinction of UV photons,
we use the empirical relationship
\begin{equation} 
	\frac{A_{V}}{N_{\mathrm{H}}}=5.3\times10^{-22} \mathrm{mag} \text{ cm}^{2}
\end{equation}
which is caused mainly by dust \citep{Draine2011}.  For reference, the typical
visual extinction $A_V$ is 8 mag in a dense molecular cloud and could increase
to 15 mag in dense cores \citep{Tielens2010}. In our fiducial model, we set
$N_{\ce{H}}=10^{22.5}$ cm$^{-2}$ and the corresponding $A_V$ is $16.8$.
We also experiment with $N_{\ce{H}}=10^{22}$ and $10^{23}$ cm$^{-2}$
for comparison and the results are shown in \S\ref{sec:Av}. 

We also note that in our fiducial model the initial chemical abundance is out of
equilibrium, in the sense that without X-ray irradiation the abundance would
still evolve relatively quickly, on a timescale of $10^6-10^7$ years.
Nevertheless, we choose this initial condition because
it agrees better with the observed abundance in several molecular clouds
\citep[e.g.][]{Wakelam2008,Quan2007}. For old molecular clouds ($\sim10^7$ years),
the chemical abundance could significantly deviate from our initial condition.
To account for this possibility, we run additional simulations in which we
evolve our network for $10^7$ years before we turn on the X-ray irradiation.  
The results are presented in \S\ref{sec:age}.

\section{Results} \label{sec:Results}

\subsection{Fiducial Model} \label{sec:Default}

\begin{figure*} \centering
\includegraphics[width=\linewidth]{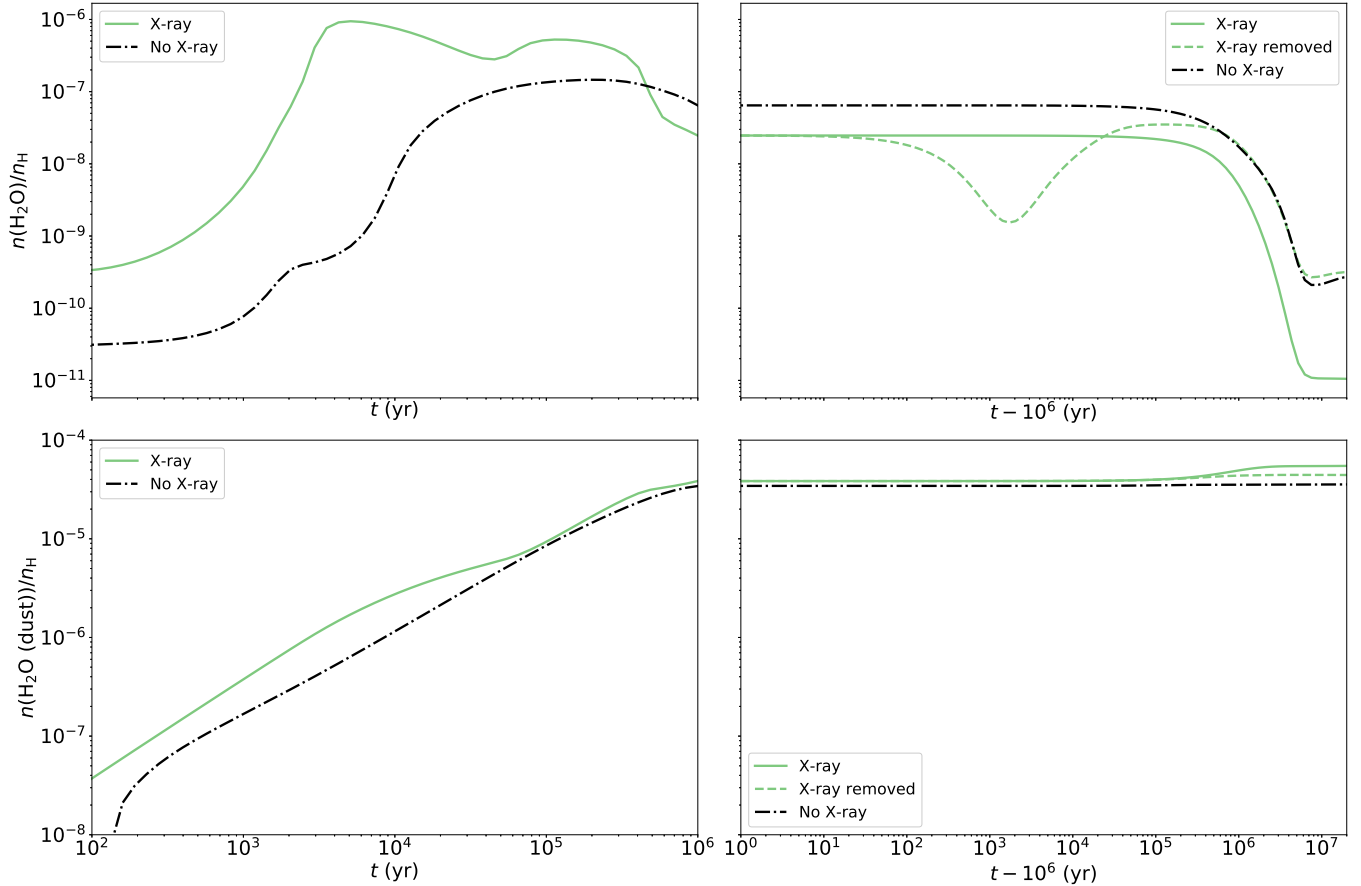} \caption{Abundances of
	\ce{H2O} in a molecular cloud $4$ kpc away from the galactic center
	and with a hydrogen column density of $N_{\ce{H}}=10^{22.5}$ cm$^{-2}$.
	The upper panels show the abundance of \ce{H2O} in the gas phase and
	the lower ones show that at the grain surface. The left panels show the
	first $10^6$ years of the chemical evolution and the right ones show
	the later $10^7$ years. The green solid curves correspond to the model
	in which the AGN persists, while the green dashed ones correspond to the
	situation where the AGN is turned off after $10^6$ years. For
	comparison, the results when there is no AGN irradiation is shown as
the black dot-dashed curves.  } \label{fig:default} \end{figure*}

\begin{figure*} \centering
	\includegraphics[width=\linewidth]{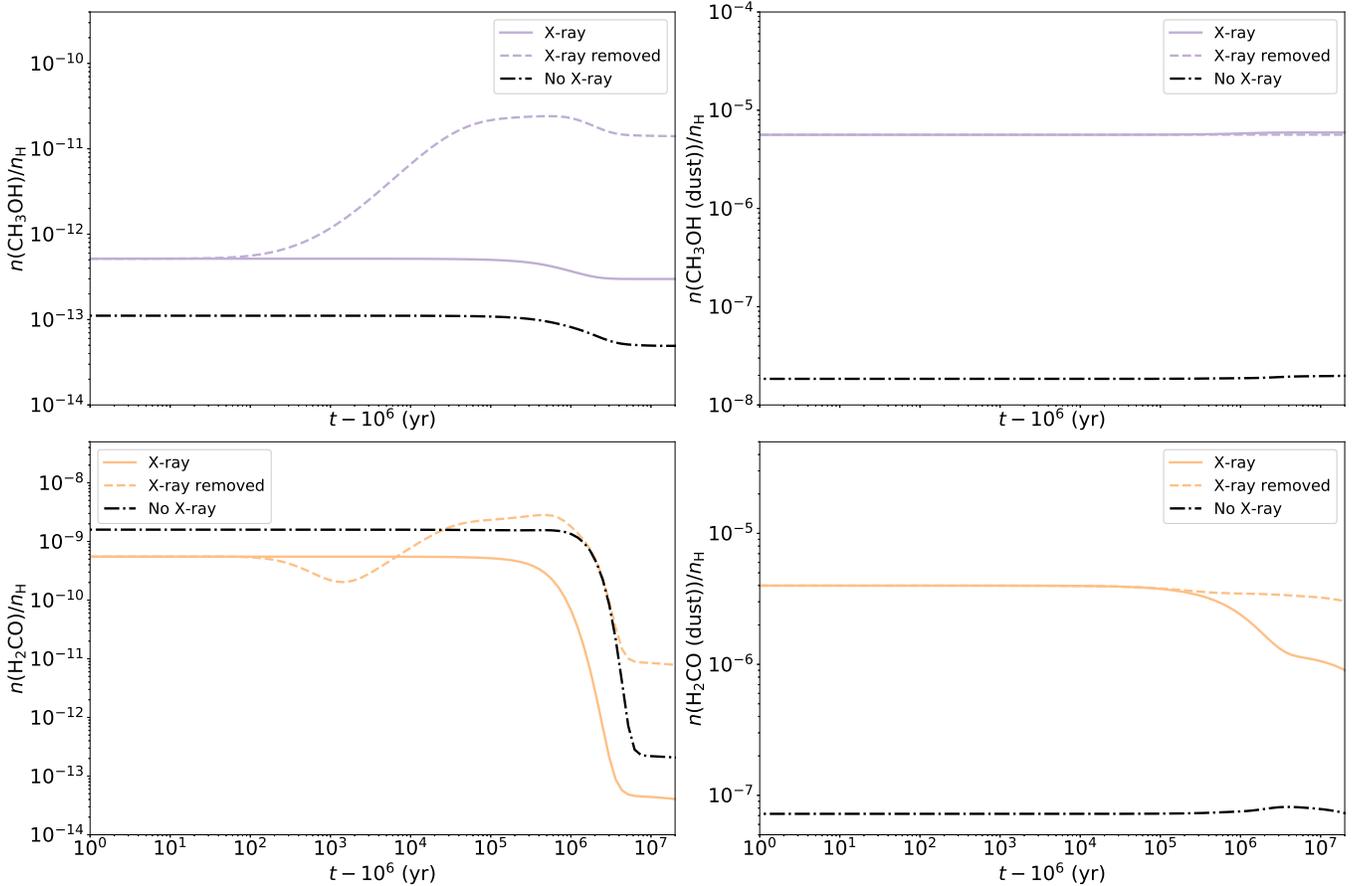}
	\caption{Evolution of the abundances of \ce{CH3OH} (upper) and
	\ce{H2CO} (lower) in the gas phase (left) and on the grain surface
	(right). The line styles have the same meanings as in
Fig.~\ref{fig:default}.} \label{fig:otherprebiotics} \end{figure*}

In our fiducial model, the Eddington ratio of the AGN is $\eta=1$.  The
molecular cloud is at a Galactic distance of $4$ kpc and has a column density
of $N_{\ce{H}}=10^{22.5}$ cm$^{-2}$ ($A_V=16.8$). The results for the chemical
evolution are shown in Figures \ref{fig:default} and
\ref{fig:otherprebiotics}. In general, the abundances of molecules on the dust
grain surface are several orders of magnitude higher than their gas-phase
counterparts. This is mainly because at low temperatures ($\sim10$ K) molecules
stay on the surface of dust grains, and the grain surface acts as a reaction
container and a catalyst leading to much more efficient formation of \ce{H2O},
\ce{CH3OH}, etc.

For \ce{H2O} (Fig.~\ref{fig:default}), on the grain surface (lower panels),
even without X-rays (the black dot-dashed curves) the abundance keeps growing
during the first $10^5-10^6$ years and afterwards saturates.  The increase is
caused mainly by two surface reactions, (i) \ce{H2 + OH -> H2O + H} and (ii)
\ce{H + OH -> H2O}.  The first reaction usually predominates because of the
high abundance of \ce{H2}, even though the activation energy is high
\citep{Hasegawa1992}. The second one has a much lower energy barrier but the
rate is limited by the low concentration of atomic \ce{H} at the grain
surface. The saturation of the water surface abundance in the late stage is due
to a balance between the photo-evaporation and the formation processes.

If we start X-ray irradiation at the very beginning (green solid curves), the
\ce{H2O} surface abundance increases faster and in the end slightly exceeds
that in the case without X-ray.  We find that the first reaction is not
significantly affected by the X-ray irradiation, because  the \ce{H2} surface
abundance only slightly decreases. On the contrary, the
second reaction becomes faster because the atomic hydrogen abundance increases
significantly due to the X-ray dissociation of \ce{H2}.  After we remove the
X-ray irradiation at the end of $10^6$ years (green dashed curve), the surface
abundance of \ce{H2O} stays more or less constant.  The value is slightly
lower than that in the case of X-ray irradiation because atomic \ce{H}
quickly recombine to form \ce{H2}, thus the rate of the second reaction
drops.

For the gas phase (the upper panels of Fig.~\ref{fig:default}),  when there is
no X-ray, the abundance of \ce{H2O} first increases until a plateau is
reached around the time of $10^4-10^5$ years. During this period, ionization
leads to the formation of \ce{H3+}, which is a highly reactive species that
can quickly produce \ce{OH+}, \ce{H2O+}, and \ce{H3O+}
\citep{Krolik1983}.  Gaseous water is synthesized mainly by the recombination
of \ce{H3O+} with electrons.  Around the time of $10^6$ years, the abundance
of gas \ce{H2O} starts to decrease because the materials such as \ce{OH+},
\ce{H2O+}, and \ce{H3O+} run out following the exhaustion of \ce{O}, so
that the formation of \ce{H2O} slows down. The final gaseous abundance is
limited mainly by the interaction with atomic cations, such as \ce{Si+}
through the ion-neutral reaction \ce{Si+ + H2O -> HSiO+ + H}.

When X-ray irradiation is turned on, the behavior of the water abundance in the
gas phase in similar to that on the grain surface. However, both the rise and
fall of the \ce{H2O} abundance are more prominent, which leads to a higher
plateau around $3\times10^5$ yr and a lower abundance at about $\gtrsim 10^6$
yr. The sharper evolution is caused by the higher cation density during the
X-ray ionization, such as \ce{H3+,\ OH+,\ H2O+}, during the early formation
stage of \ce{H2O}, as well as the higher abundance of \ce{Si+} during the
later stage of \ce{H2O} destruction. After we remove the X-rays, the
abundance of \ce{H2O} first decreases but within about $10^5$ years recovers
the value which is seen in the case of no X-ray irradiation.  The recovery is
caused by the dissociative recombination, \ce{H3O+ + e- -> H2O + H}, as well
as the photo-evaporation from the grain surface.

For \ce{CH3OH}, the evolution of the abundance after the first $10^6$ yr is
shown in the two upper panels of Figure~\ref{fig:otherprebiotics}.  We do
not show the evolution in the first $10^6$ years because it is a monotonic
increase. Also, we are more interested in the molecular abundances millions of
years after the AGN turns off. At the grain surface (right panel), when there
is no X-ray irradiation, the abundance is almost constant over $10^7$ years of
evolution.  The dominant synthesis process is hydrogenation, via \ce{CO -> HOC
-> CHOH -> CH2OH -> CH3OH}.  When X-ray irradiation is turned on, the abundance
of \ce{CH3OH} becomes more than two orders of magnitude higher. The increase is
caused by a faster hydrogenation process, as the result of a larger
atomic-hydrogen abundance due to the X-ray disassociation.  Even after we
remove the X-ray, the \ce{CH3OH} surface abundance stays nearly the same.  This
is because in our model there is no chemical reaction at the grain surface for
\ce{CH3OH} destruction \citep[also see][]{Hasegawa1992} and also the
photo-evaporation of \ce{CH3OH} is inefficient.

In the gas phase (upper-left panel), without X-ray irradiation, the abundance
of \ce{CH3OH} stays constant for about $10^6$ years and afterwards slightly
declines.  Gaseous \ce{CH3OH} mainly comes from the grain surface, and the
later decline is caused by the reactions with cations, e.g., \ce{CH3OH + He+
-> OH+ + CH3 + He / OH + CH3+ + He}.  With X-rays and for the particular
Galactic distance of our choice, i.e., $4$ kpc, the abundance of  \ce{CH3OH}
increases by about one order of magnitude relative to that in the case without
X-rays. The increase is closely related to the enhancement of the \ce{CH3OH}
abundance on dust grains.  After we remove the X-rays, the abundance increases
even more because cations recombine and the destruction of \ce{CH3OH} slows
down.

For \ce{H2CO}, the results are shown in the lower two panels of
Figure~\ref{fig:otherprebiotics}. Unlike \ce{CH3OH} which forms only at the
grain surface, \ce{H2CO} can form both on the surface of grains, mainly
through hydrogenation \ce{CO -> HCO -> H2CO}, and in gas phase, via \ce{CH3
+ O -> H2CO + H}. Moreover, There is also one pathway for \ce{H2CO} on the
grain surface to dissociate, \ce{H2CO + H -> HCO + H2}. The \ce{O} atom can
then be transferred to species like \ce{CO2} through reaction \ce{O + HCO ->
CO2 + H}. Although the activation energy of this dissociation is high, the
reaction rate is still orders of magnitude higher than photo-evaporation. 

With these differences in mind, we can understand the behavior of \ce{H2CO}
at the grain surface (lower-right panel).  During the X-ray irradiation, the
formation rate is first enhanced due to a higher hydrogenation rate.  In the
last several million years of our simulation the abundance of \ce{H2CO}
significantly decreases, because the disassociation process becomes important and \ce{O} atoms are slowly transferred to species like \ce{CO2}.
After we remove the X-ray irradiation, the decrease of the surface abundance of
\ce{H2CO} becomes slower because the surface abundance of \ce{H} falls
substantially. Because the formation (hydrogenation) of \ce{H2CO} is also
proportional to the surface abundance of \ce{H}, we do not see an increase of
\ce{H2CO} after we remove the X-ray irradiation.

In the gas (lower-left panel), even without X-rays, the abundance of
\ce{H2CO} drops by almost four orders of magnitude by the end of the
$10^6-10^7$ years of evolution.  The drop is mainly caused by the exhaustion of
\ce{CH3} and \ce{O}, so that the rate of the reaction \ce{CH3 + O -> H2CO
+ H} significantly decrease by the end of our simulation.  With X-rays, the
abundance decreases relative to that in the case without X-ray irradiation, by
a factor of a few. The drop is due to the higher abundance of cations. In this
case, the destructive reactions, such as \ce{H2CO + S+ -> HCO+ + HS}, become
more efficient.  When we remove the X-rays, the \ce{H2CO} abundance first
recovers the value in the case of no X-ray irradiation and in about $10^6$
years significantly decreases due to the exhaustion of \ce{CH3}. Finally, the
abundance balances at a value which is higher than that in the no-X-ray case,
because more \ce{CH3} has been produced during the episode of X-ray
irradiation.
	
For completeness, we show in the \ref{sec:appendix} the evolution of
the other species that are often detected in molecular clouds. In general, we
also find that their abundances with and without X-rays are different.

\subsection{Distribution in the Galactic Disk} \label{sec:Distribution}

\begin{figure*} \centering
	\includegraphics[width=\linewidth]{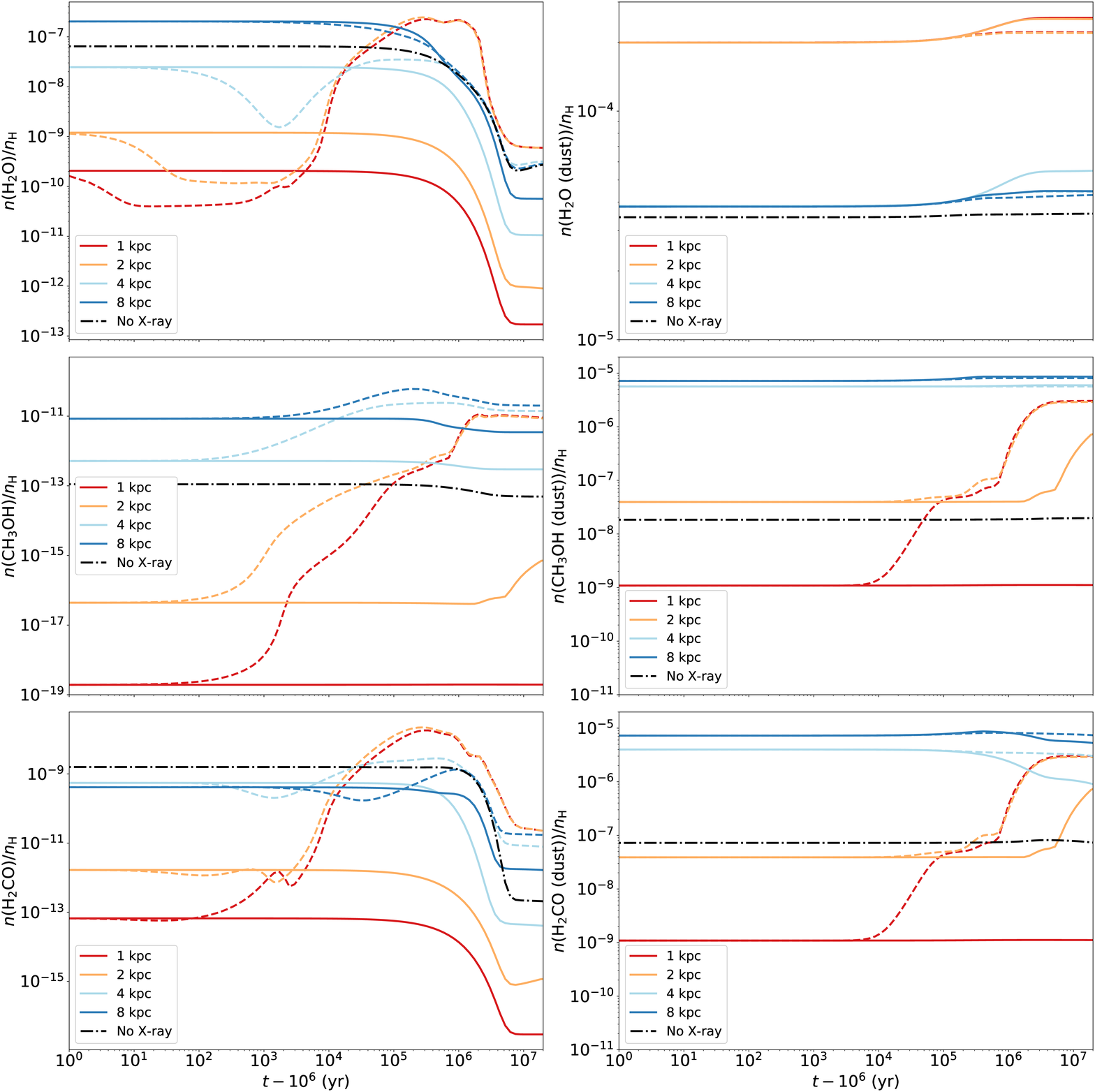}
	\caption{Evolution of the abundances of \species\space (from top to
	bottom) inside the molecular clouds at different Galactic distances
	($1$, $2$, $4$, and $8$ kpc).  The left panels show the abundances in
the gas phase and the right ones show the surface abundances. The lines styles
have the same meanings as in in Fig.~\ref{fig:default}.} \label{fig:galactic}
\end{figure*}

Since the X-ray flux decreases with increasing distance from the Galactic
Center, we also explore the chemical evolution at different Galactic distances.
The results are shown in Figure \ref{fig:galactic}, where the column density of
each cloud is still set to $N_{\ce{H}}=10^{22.5}$ cm$^{-2}$.

On the grain surface (right panels), when there is X-ray irradiation (solid
curves), the abundance of \ce{H2O} keeps increasing as the Galactic distance
decreases.  The increase is caused by the higher abundance of atomic
hydrogen as the ionization flux intensifies. For \ce{CH3OH} and
\ce{H2CO}, however, the surface abundance maximizes at a Galactic distance of
$4-8$ kpc, and further increasing or decreasing the distance would both lead to a
lower abundance. The location where these two abundances peak is determined by
two competing processes. On one hand, an important precursor of both species,
\ce{CO}, can be destroyed by X-rays and hence cannot exist too close to the
Galactic Center when the AGN is on. On the other, the abundance of \ce{H}
increases towards the AGN due to X-ray disassociation. These two processes
balance at about $4-8$ kpc and produce the highest abundance of \ce{CH3OH} and
\ce{H2CO} there.

To understand the behavior of \species after we turn off the X-ray irradiation
(dashed lines), we should first understand the evolution of \ce{H2} and
\ce{CO}.  The major difference between the two species is that while the
\ce{H2} abundance is lowered by several percents during the X-ray
irradiation, the abundance of \ce{CO} could be lowered by orders of magnitude
depending on the Galactic distance.  Therefore, after we turn off the X-rays,
the abundance of \ce{H2} only slightly increases in percentage, due to recombination, and
recovers the equilibrium in the case of no X-rays, but the abundance of
\ce{CO} increases more drastically.  The difference in the recovery rate
causes the different behavior of \ce{H2O} versus \ce{CH3OH} and \ce{H2CO}
at the grain surface. We can see that after we turn off X-rays, the surface
abundance of \ce{H2O} increases slightly, because there is slightly more
\ce{H2} for the reaction \ce{H2 + OH -> H2O + H}. For \ce{CH3OH} and
\ce{H2CO}, however, there is a lot more \ce{CO} produced within a short
period of time, especially at small Galactic distances. For example,  at $1$
kpc the abundances of \ce{CH3OH} and \ce{H2CO} rise by orders of magnitude
after the X-ray irradiation is turned off. One interesting result, which is
shared by all the three molecular species (\species) and at all the simulated
Galactic distances, is that the final surface abundance after we remove X-rays
is higher than the abundance in the case without X-ray irradiation. 

In the gas phase (left panels), the behavior of \species also depends on the
Galactic distance. For \ce{H2O}, the X-ray irradiation in general suppresses
the abundance within a timescale of $10^7$ years, because, as we have explained
in \S\ref{sec:Default}, destructive cations form in large amount but water
formation on the grain surface is relatively inefficient. Moreover, the final
abundance decreases with decreasing distance, because the X-ray flux is higher
at smaller distance.  For \ce{CH3OH} and \ce{H2CO}, destructive cations
also form due to the X-ray irradiation.  That is why at small Galactic
distances such as $1$ and $2$ kpc, the abundances of \ce{CH3OH} and
\ce{H2CO} are suppressed when the AGN is on.  However, at large distances
such as $4$ and $8$ kpc, the abundances are not significantly suppressed and
could even exceed those in the cases without X-rays. The reason is that at
these distances the X-ray flux is not as strong so that  \ce{CH3OH} and
\ce{H2CO} form very fast on the grain surface, which, through
photo-evaporation, could also enrich the gas.

After we turn off the X-ray irradiation, the gaseous abundances of \species all
increase within a timescale of $10^7$ years, regardless of the Galactic
distance. The cause is mainly the recombination of cations, which in turn
lowers the destruction rate. Photo-evaporation also contributes considerably.
Even for \ce{H2O}, whose gaseous concentration is not as sensitive to the
surface abundance, photo-evaporation enriches the gas species significantly
after $\sim5\times10^6$ years, when the dissociative recombination has
significantly slowed down. We find that at all distances, after we remove the
X-rays, the final abundances of gaseous \species exceed those in the case
without an AGN or with a constant AGN.
	
These results suggest that a relatively short episode ($10^6$ years) of X-ray
irradiation could indeed, in the following $10^7$ years, leave an imprint in
the molecular abundances throughout the MW.

\subsection{Distribution Inside Molecular Cloud} \label{sec:Av} 

\begin{figure*} \centering
	\includegraphics[width=\linewidth]{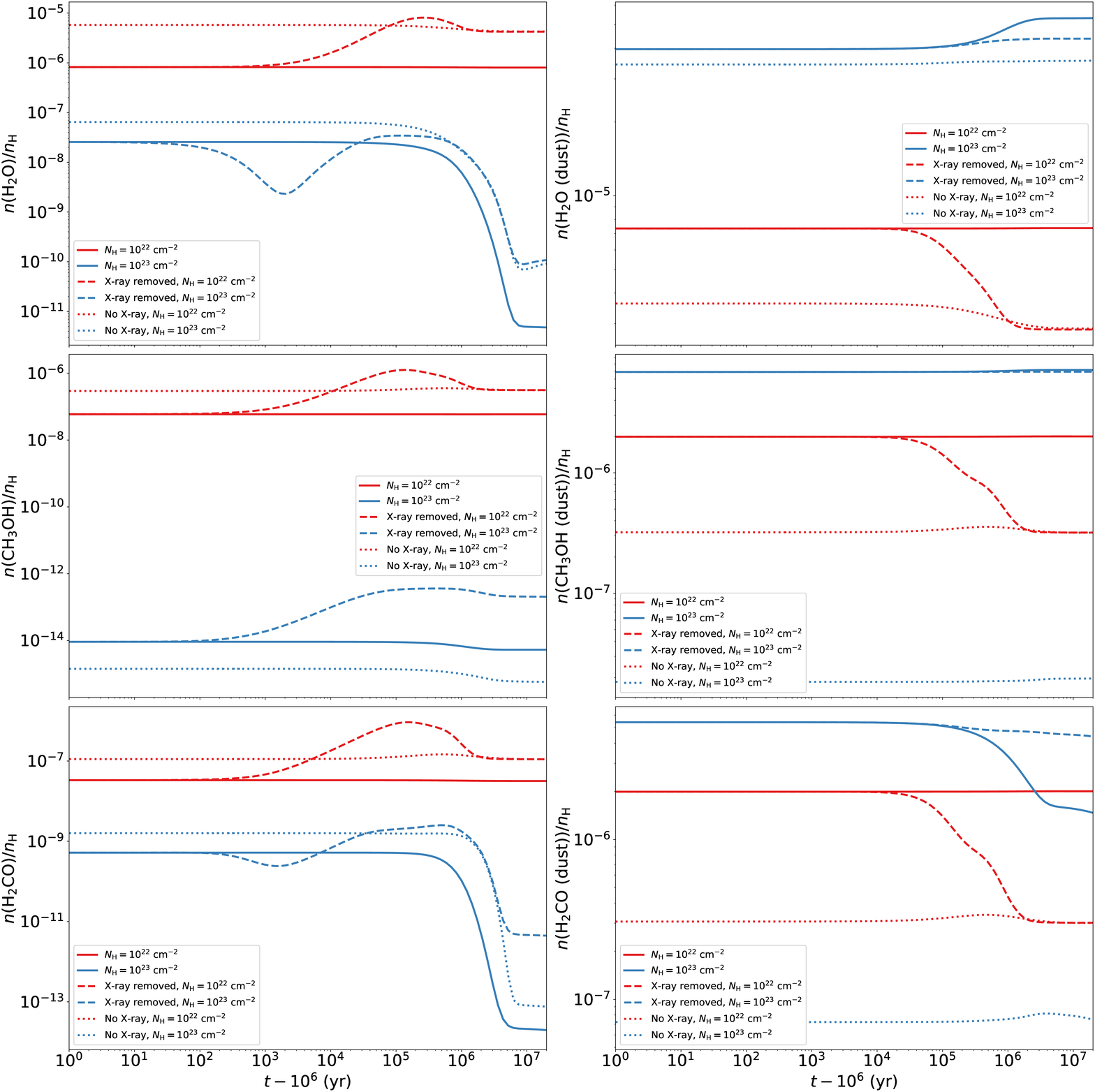}
	\caption{Abundances of \species (from top to bottom) at different depth
	of a molecular cloud, located at $4$ kpc from the Galactic Center.  The
	depth is proportional to the column density $N_{\ce{H}}$.  The red
	curves correspond to a column density of $N_{\ce{H}}=10^{22}\,{\rm
	cm^{-2}}$ and the blue ones to $10^{23}\,{\rm cm}^{-2}$.  The left
	panels show the abundances in the gas phase and the right ones show the
	abundances on the grain surface.  } \label{fig:column} \end{figure*}

To understand the distribution of \species as a function of the depth inside
a molecular cloud, we compute the chemical evolution assuming different
column densities ($N_{\ce{H}}$) for the cloud. The results are shown in 
Figure~\ref{fig:column}. In these calculations, the Galactic distance
of the cloud is fixed at $4$ kpc.

We find that when the column density is high, i.e., $N_{\ce{H}}=10^{23} {\rm
cm}^{-2}$ (or equivalently $A_V=53$), the abundances (see the blue curves) are
slightly, but systematically lower than those in the fiducial model. This trend
can be seen both in the gas phase and on the grain surface.  The main cause of
this systematic change is the lower UV flux due to higher extinction.  The hard
X-ray flux from the AGN is only slightly attenuated inside the molecular cloud,
and hence when we turn on and off the AGN the evolution of the three species is
similar to the evolution in our fiducial model.

In low-extinction regions ($N_{\ce{H}}=10^{22}$ cm$^{-2}$, $A_V=5.3$, the red
curves), the surface abundances (right panels) are more significantly affected
by the rise of the UV irradiation.  Even when there is no X-ray radiation
(dotted curves), the \ce{H2O} abundance is lower than that in the fiducial
case by one order of magnitude. On the contrary, the surface abundances of
\ce{H2CO} and \ce{CH3OH} are higher than those in the fiducial model.  This
result is caused by two competing processes.  On one hand, surface species
evaporate more rapidly in the presence of a higher UV flux. On the other, the
higher density of \ce{H} due to UV ionization/dissociation enhances the rate
of hydrogenation.  Since the synthesis of \ce{H2O} uses mainly \ce{H2}, not
\ce{H}, a higher UV flux would increase the evaporation rate more than the
hydrogenation one, causing the surface abundance of water to decrease. For
\ce{CH3OH} and \ce{H2CO}, the hydrogenation rate is relatively higher than
the evaporation rate, so that the surface abundances increase when the UV flux
becomes higher. 

If we turn on X-rays, the surface abundances of all the three species increase
(compare the red-dotted and the red-solid curves), because of an enhancement of
the \ce{H} density on the grain surface.  However, compared to those in the
high-extinction regions (blue solid curves), the surface abundances in the
low-extinction regions is reduced, because of a higher evaporation rate induced
by X-rays in the latter case.  After we remove the X-ray irradiation, within
$\sim10^6$  years the abundances recover the values in the case of no X-rays.
The quick recovery is closely correlated with the high reaction and evaporation
rate in the low-extinction regions.

In the gas phase of the low-extinction regions (red curves in the left panels),
the abundances of \species without X-rays are much higher than those in the
high-extinction ones. After we turn on the X-ray irradiation, the abundances
become lower, but are still higher than those in the high-extinction regions.
The high abundances in the gas are caused mainly by the high photo-evaporation
rate on the grain surface. After we turn off X-rays, the gas abundances quickly
recover the values in the no-X-ray case, and hence the imprint of the X-rays
disappears within about $10^6$ years.
	
\subsection{Age of Molecular Cloud}\label{sec:age} 

\begin{figure*} \centering
\includegraphics[width=\linewidth]{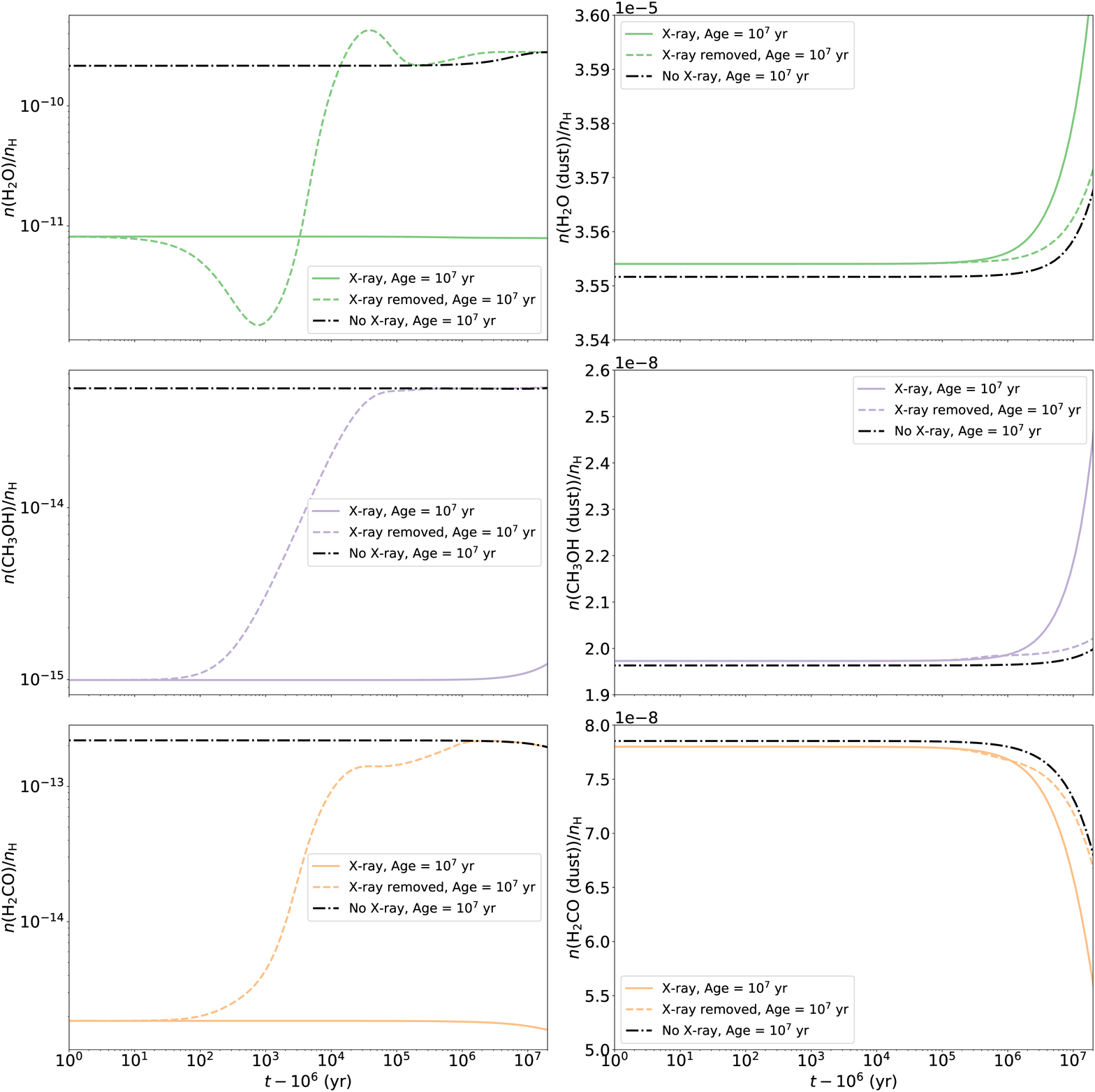} \caption{
The same as in our fiducial model but the molecular clouds are 
$10^7$ years old in the initial condition of our simulation. 
} \label{fig:age} 
\end{figure*}

To see the response of a more evolved molecular cloud to the X-rays from the
AGN, we run our chemical-reaction network without X-rays for $10^7$ years and
take the resulting chemical abundances as our new initial condition
for the later simulations. The other parameters are the same as in the
fiducial model.  The results are presented in Figure~\ref{fig:age}.
We see that the abundances in the case without X-rays no longer significantly
vary with time, both in the gas and on the grain surface, suggesting that the
chemical reactions have reached an equilibrium on a timescale of $10^7$ years.

If we turn on X-ray irradiation after the $10^7$ years of isolated evolution,
on the grain surface (right panels) the abundances of \ce{H2O} and
\ce{CH3OH} will increase, as we have seen in the fiducial model, but the
abundance of \ce{H2CO} decreases. The decrease is caused by the reaction
\ce{H2CO + H -> HCO + H2}, which predominates in the current conditions.  We
notice that the variation of the abundances of \ce{CH3OH} and \ce{H2CO}
after we turn on X-rays is not as prominent as that in a less evolved cloud
(see our fiducial model). This difference indicates that a chemically more
evolved cloud is more resilient to X-ray irradiation.  After we remove the
X-rays, the surface abundances return to the values in the case of no X-ray
radiation, on a timescale of about $10^7$ years.

In the gas phase (left panels), the X-ray irradiation significantly lowers the
abundances of all the three molecules. This behavior is different from that in
a less-evolved could, where the \ce{CH3OH} abundance is enhanced by X-rays
(see Fig.~\ref{fig:otherprebiotics}). After we remove the X-rays, the
abundances rise more quickly than that in a less-evolved cloud. As a result of
the quick rise, a recovery of the equilibrium state for no X-rays is seen at
about $10^4-10^5$ years. After this time, the imprint of an AGN is lost. 

\section{Diagnostics}\label{sec:diag}

\begin{figure*} 
\centering \includegraphics[width=\linewidth]{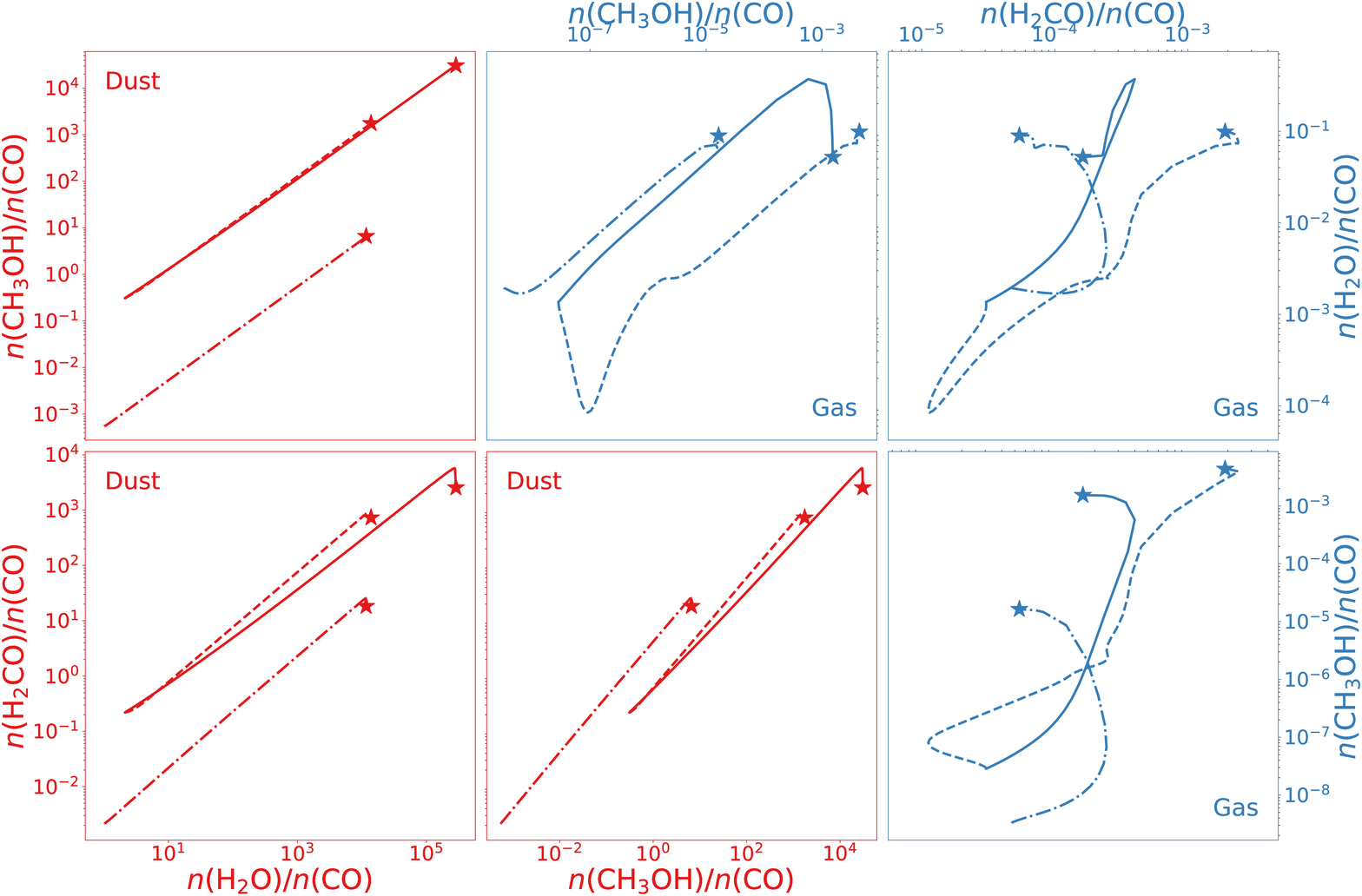}
	\caption{Diagnostics of the effect of X-ray irradiation on the
	molecular abundances. The abundances are calculated relative to the gas \ce{CO} for easier comparison
	with observations. The three blue panels labeled with ``Gas''
	refer to the abundances in the gas phase, and the red ones 
	with ``Dust''refer to the surface abundances.  The curves correspond to the results from our
	fiducial model shown in Fig.s~\ref{fig:default} and
\ref{fig:otherprebiotics}. The line styles have the same meanings, i.e.,
the dot-dashed curves correspond to the evolution without X-rays,
the solid ones correspond to the evolution with constant X-ray irradiation,
and the dashed ones show the evolution when the X-ray is turned off after
an initial $10^6$ years of irradiation. 
	The
stars mark the end of our simulations, at the time of $5\times10^7$ years.   }
\label{fig:diag} 
\end{figure*}

Here we provide possible diagnostics for future observations by showing
in Figure~\ref{fig:diag} the abundance ratios of \species with respect to
\ce{CO}.  We choose \ce{CO} as the reference because it is the second most
abundant gas molecule (after \ce{H2}) and relatively easy to detect in
molecular clouds.  Molecules on the grain surface are not directly detectable,
but we plot their abundance ratios relative to the gas-phase \ce{CO} as well,
because surface molecules could be rapidly released into the gas in certain
processes which are not considered in this work, for example, by turbulence or
shock \citep{federman91,Garcia-Burillo10,Harada2018}. We show only the results
from our fiducial model. The other models generally lead to the same
conclusion.

The three blue panels labeled with ``Gas'' in Figure~\ref{fig:diag} show
the abundance ratios in the gas phase. From these diagrams, we find that  by
the end of our simulations (marked by the star symbols), the locations of the
molecular clouds exposed to X-ray irradiation (solid and dashed curves) are
well separated from those clouds without X-ray irradiation (dot-dashed curves).
The divergence of the curves starts the earliest in the diagram of
$\ce{H2CO/CO}$ against $\ce{CH3OH/CO}$. It is worth noting that at the column
densities of our interest, the low-J transitions of \ce{H2CO} and \ce{CH3OH}
are optically thin, therefore they are good tracers of the past activities of
Sgr A*.

The three red panels labeled with ``Dust'' show the abundances of the
\species on the grain surface relative to the gaseous abundance of \ce{CO}. The
evolutionary tracks with and without X-rays, as well as the final locations of
the molecular clouds,  are well separated in these diagrams. This result
indicates that the molecular clouds showing strong signs of turbulence or shock
could provide better diagnostics of the activity of Sgr A*, because the surface
molecules could have been released into the gas.

For the other neutral species simulated in this work, their
diagnostics are shown in \ref{sec:appendix}.

\section{Caveats}
\label{sec:caveat} 

Besides photo-desorption (see \S\ref{sec:grain}), X-rays could also excite an
electron to a sufficiently high energy so that it escapes from the grain
surface \citep{Draine2011}. These photoelectrons from dust grains, in
principle, are energetic enough to ionize the gas molecules, but we did not
consider them in this work because of the uncertainty in the calculation of
their number \citep{Bakes1994,Weingartner2001,Weingartner2006}.  The following
estimation shows that by neglecting these photoelectrons, we could have
underestimated the ionization rate of molecular gas by a factor of $2-9$, and
hence underestimated the effect of the AGN on the molecules in the gas phase.

Following the work of \cite{Weingartner2006} who modeled photoelectric emission
from the grains exposed to extreme ultraviolet and X-ray radiation, we assume
that (i) the typical dust particles are carbonaceous grains with a radius of
$0.1 \mu$m (cross section $\sim3\times10^{-10}$ cm$^2$), (ii) after each
absorption of a hard-X-ray photon ($h\nu>1$ keV) the probability of a
photoelectron (either primary or secondary) escaping the bulk solid is $0.1-1$,
and (iii) the photoionization cross section of hydrogen atom in ground state is
$10^{-23}$ cm$^2$ at 1 keV. Moreover, we adopt the previous assumptions of a
gas-to-dust ratio of $100$ (in mass) and a typical grain density of $3$
g/cm$^3$. We find that the ratio between the number density of hydrogen atoms
and that of carbonaceous grains is $8\times10^{11}$. Given these numbers, we
find that for each X-ray photon of $1$ keV the average yield of photoelectrons
from dust grains is about $24-240$.  The exact number depends on the sizes of
grains which is not well constrained by observations. However, it is about
$0.8-8$ times the yield of photoelectrons from gaseous hydrogen (see
\S\ref{sec:X-ray}). Therefore, the effect of the photoelectrons
from dust grains deserves further investigation. 

The loss of photoelectrons would leave the dust grains positively charged.  The
X-ray induced grain charging has not been modelled in detail until recently
(e.g. \citealp{lb2019}). Charged grains, just like ambient electrons, can
affect the temperature of free electrons through Coulomb scattering. They may
also react with gaseous anions. However, both free electrons and gaseous anions
are rare in the system of our interest. Therefore, we do not consider grain
charging in this work.

\section{Summary and Conclusion}
\label{sec:Summary} 

Motivated by the observational evidence that the SMBH in the Galactic Center
could be an AGN several million years ago, we studied the impact of the
corresponding X-ray irradiation on the molecular chemistry in the MW.  Our main
results are summarized as follows.

\begin{enumerate}

\item In our fiducial model, a molecular cloud is located at $4$ kpc from
	the Galactic Center and has a column density of $10^{22.5}\,\rm{cm}^{-2}$. 
	The X-ray irradiation from the AGN
		could slightly enhance the abundance of \ce{H2O} on the
		surface of dust grains and, at the same time, suppress the
		water abundance in the gas phase. After the AGN turns off, in
		the following $10^7$ years, the abundance on the grain surfaces
		remains slightly higher than that in the case without X-rays,
		while the abundance in the gas phase almost recovers the value of the
		no-X-ray case.

\item For \ce{CH3OH} and \ce{H2CO}, our fiducial model shows that the
	abundances on the grain surface could be enhanced by one to two orders
		of magnitude during the X-ray irradiation. The enhancement
		could sustain for about $10^7$ years even after the AGN is
		turned off. In the gas phase, the abundance of \ce{CH3OH}
		is enhanced during the AGN episode by about one order of
                magnitude, but that of \ce{H2CO} is reduced slightly. 
		Interestingly, after we turn off the X-ray irradiation, both 
		abundances rise significantly. Therefore, by the end of the 
		$10^7$ years of simulation, the final gas abundances are higher
		than the values in the no-X-ray case, by about two orders of magnitude.
	
\item The exact values of the molecular abundances during and after the AGN
	episode depend on the distance of the molecular cloud from the Galactic
		Center. However, there is one feature which appears to be
		common at different Galactic distances.  For a cloud similar to
		that in our fiducial model, if we irradiate it with X-ray for
		$10^6$ years and then turn off the irradiation for $10^7$
		years, the final gas and the surface abundances are higher than
		the case with no X-ray irradiation. The enhancement is the most
		prominent at small Galactic distances, such as $\la2$ kpc.

\item After the AGN turns off, the recovery of the molecular abundances is
	faster in low-extinction regions ($N_H \la10^{22}\,\rm{cm}^{-2} $), such
		as the surface of a molecular cloud or a low-column-density cloud.
		Therefore, the imprint of a past AGN should be easier to
		observe in high-column-density regions, e.g., $N_H \ga10^{22.5}\,\rm{cm}^{-2} $.

\item Older molecular clouds (e.g. $10^7$ years) recover more rapidly after the
	AGN turns off. Therefore, young molecular clouds are more likely to
		bear the chemical imprint of a past AGN.
	
\end{enumerate}

These results suggest that the abundances of molecular species in the MW could be
significantly affected and reach a new equilibrium during the past AGN
activities of Sgr A*. The chemical imprint of the most recent AGN, which could
have occurred several million years ago, may still be found today in those
young, high-density molecular clouds residing at relatively small distances from
the Galactic Center.
	
\acknowledgments
	
This project is under the framework of the Undergraduate Research and Training
Program of Peking University and is sponsored by the National Innovation
Training Program and School of Physics, Peking University.  We thank Jieying
Liu for kindly providing us with the AGN spectra of her numerical simulations
and Zhu Liu for his help with  the UV/X-ray extinction. We also thank Ke Wang,
Guangshuai Zhang, Emma Yu, and Sarah Dodson-Robinson for compiling a reading
list of the relevant literatures.  X.C.  acknowledges the supported by the NSFC
grant Nos. 11721303 and 11991053.  F.D. is supported by the Hundred-Talent
Program (Chinese Academy of Sciences) through grant 2017-089 and by grants from
NSFC (No. 11873094 and 11873097).

\appendix

\section{Surface Reaction Network}\label{sec:surface_reaction}

Table~\ref{table:Grain_Reaction} shows surface reactions in our model selected 
from the network in \cite{Hasegawa1992}. The last column shows the reaction barrier.

\startlongtable \begin{deluxetable}{ccccc} \tablecaption{Surface
		reaction network.\label{table:Grain_Reaction}} \startdata
	\tablehead{Reactant 1 &            Reactant 2 &
		Product 1 &         Product 2 & $E_a$ (K)} 
	\ce{H} & \ce{H} & \ce{H2} & & 0 \\
	\ce{H} & \ce{C} & \ce{CH} & & 0\\
	\ce{H} & \ce{N} & \ce{NH} & & 0 \\
	\ce{H} & \ce{O} & \ce{OH} &  & 0\\
	\ce{H} & \ce{CH} & \ce{CH2} &  & 0 \\
	\ce{H} & \ce{NH} & \ce{NH2} &  & 0 \\
	\ce{H} & \ce{OH} & \ce{H2O} &  & 0 \\
	\ce{H} & \ce{C2} & \ce{C2H} &  & 0 \\
	\ce{H} & \ce{CN} &	\ce{HCN} &  & 0 \\
	\ce{H} & \ce{CO} & \ce{HOC} &  & 1000 \\
	\ce{H} & \ce{CO} & \ce{HCO} &  & 1000 \\
	\ce{H} & \ce{NO} & \ce{HNO} &  & 0 \\
	\ce{H} & \ce{O2} & \ce{O2H} &  & 1200 \\
	\ce{H} & \ce{N2} & \ce{N2H} &  & 1200 \\
	\ce{H} & \ce{CH2} & \ce{CH3} &  & 0 \\
	\ce{H} & \ce{NH2} & \ce{NH3} &  & 0 \\
	\ce{H} & \ce{C2H} & \ce{C2H2} &  & 0 \\
	\ce{H} & \ce{HOC} & \ce{CHOH} &  & 0 \\
	\ce{H} & \ce{HCO} & \ce{H2CO} &  & 0 \\
	\ce{H} & \ce{O2H} & \ce{H2O2} &  & 0 \\
	\ce{H} & \ce{O3} & \ce{O2} & \ce{OH} & 450 \\
	\ce{H} & \ce{C2N} & \ce{HCCN} &  & 0 \\
	\ce{H} & \ce{N2H} & \ce{N2H2} &  & 0 \\
	\ce{H} & \ce{CH3} & \ce{CH4} &  & 0 \\
	\ce{H} & \ce{H2CO} & \ce{HCO} & \ce{H2} & 1850 \\
	\ce{H} & \ce{CHOH} & \ce{CH2OH}&  & 0 \\
	\ce{H} & \ce{H2O2} & \ce{H2O} & \ce{OH} & 1400 \\
	\ce{H} & \ce{N2H2} & \ce{N2H} & \ce{H2} & 650 \\
	\ce{H} & \ce{HCCN} & \ce{CH2CN} &  & 0 \\
	\ce{H} & \ce{C2H2} & 	\ce{C2H3} &  & 1210 \\
	\ce{H} & \ce{C2H3} & \ce{C2H4} &  & 0 \\
	\ce{H} & \ce{CH2OH} & \ce{CH3OH} &  & 0 \\
	\ce{H} & \ce{CH2CN} & \ce{CH3CN} &  & 0 \\
	\ce{H} & \ce{C2H4} & \ce{C2H5} &  & 750 \\
	\ce{H} & \ce{C2H5} & \ce{C2H6} &  & 0 \\
	\ce{H2} & \ce{OH} & \ce{H2O} & \ce{H} & 2600 \\
	\ce{C} & \ce{C} & \ce{C2} &  & 0 \\
	\ce{C} & \ce{N} & \ce{CN} &  & 0 \\
	\ce{C} & \ce{O} & \ce{CO} &  & 0 \\
	\ce{C} & \ce{CH} & \ce{C2H} &  & 0 \\
	\ce{C} & \ce{NH} & \ce{HNC} &  & 0 \\
	\ce{C} & \ce{OH} & \ce{HOC} &  & 0 \\
	\ce{C} & \ce{OH} & \ce{CO} & \ce{H} & 0 \\
	\ce{C} & \ce{CN} & \ce{C2N} &  & 0 \\
	\ce{C} & \ce{NO} & \ce{OCN} &  & 0 \\
	\ce{C} & \ce{O2} & \ce{CO} & \ce{O} & 0 \\
	\ce{C} & \ce{CH2} & \ce{C2H2} &	 & 0 \\
	\ce{C} & \ce{NH2} & \ce{HNC} & \ce{H} & 0 \\
	\ce{C} & \ce{OCN} & \ce{CO} & \ce{CN} & 0 \\
	\ce{C} & \ce{CH3} & \ce{C2H3} &  & 0 \\
	\ce{N} & \ce{N} & \ce{N2} &  & 0 \\
	\ce{N} & \ce{O} & \ce{NO} &  & 0 \\
	\ce{N} & \ce{CH} & \ce{HCN} &  & 0 \\
	\ce{N} & \ce{NH} & \ce{N2H} &  & 0 \\
	\ce{N} & \ce{C2} & \ce{C2N} &  & 0 \\
	\ce{N} & \ce{NH2} & \ce{N2H2} &  & 0 \\
	\ce{O} & \ce{O} & \ce{O2} &  & 0 \\
	\ce{O} & \ce{CH} & \ce{HCO} &  & 0 \\
	\ce{O} & \ce{NH} & \ce{HNO} &  & 0 \\
	\ce{O} & \ce{OH} & \ce{O2H} &  & 0 \\
	\ce{O} & \ce{CN} & \ce{OCN} &  & 0 \\
	\ce{O} & \ce{O2} & 	\ce{O3} &  & 0 \\
	\ce{O} & \ce{CO} & \ce{CO2} &  & 0 \\
	\ce{O} & \ce{HCO} & \ce{CO2} & \ce{H} & 0 \\
	\ce{O} & \ce{CH2} & \ce{H2CO} &  & 0 \\
	\ce{O} & \ce{CH3} & \ce{CH2OH} &  & 0 \\
	\ce{CH} & \ce{CH} & \ce{C2H2} &  & 0 \\
	\ce{CH} & \ce{OH} & \ce{CHOH} &  & 0 \\
	\ce{CH} &\ce{HNO} & \ce{NO} & \ce{CH2} & 0 \\
	\ce{CH} & \ce{CH3} & \ce{C2H4} &  & 0 \\
	\ce{OH} & \ce{OH} & \ce{H2O2} &  & 0 \\
	\ce{OH} & 	\ce{CH2} & \ce{CH2OH} &  & 0  \enddata 
\end{deluxetable}

\section{Other Important Molecular Species} \label{sec:appendix}

\begin{figure*}[h]
		\centering
		\includegraphics[width=0.9\linewidth]{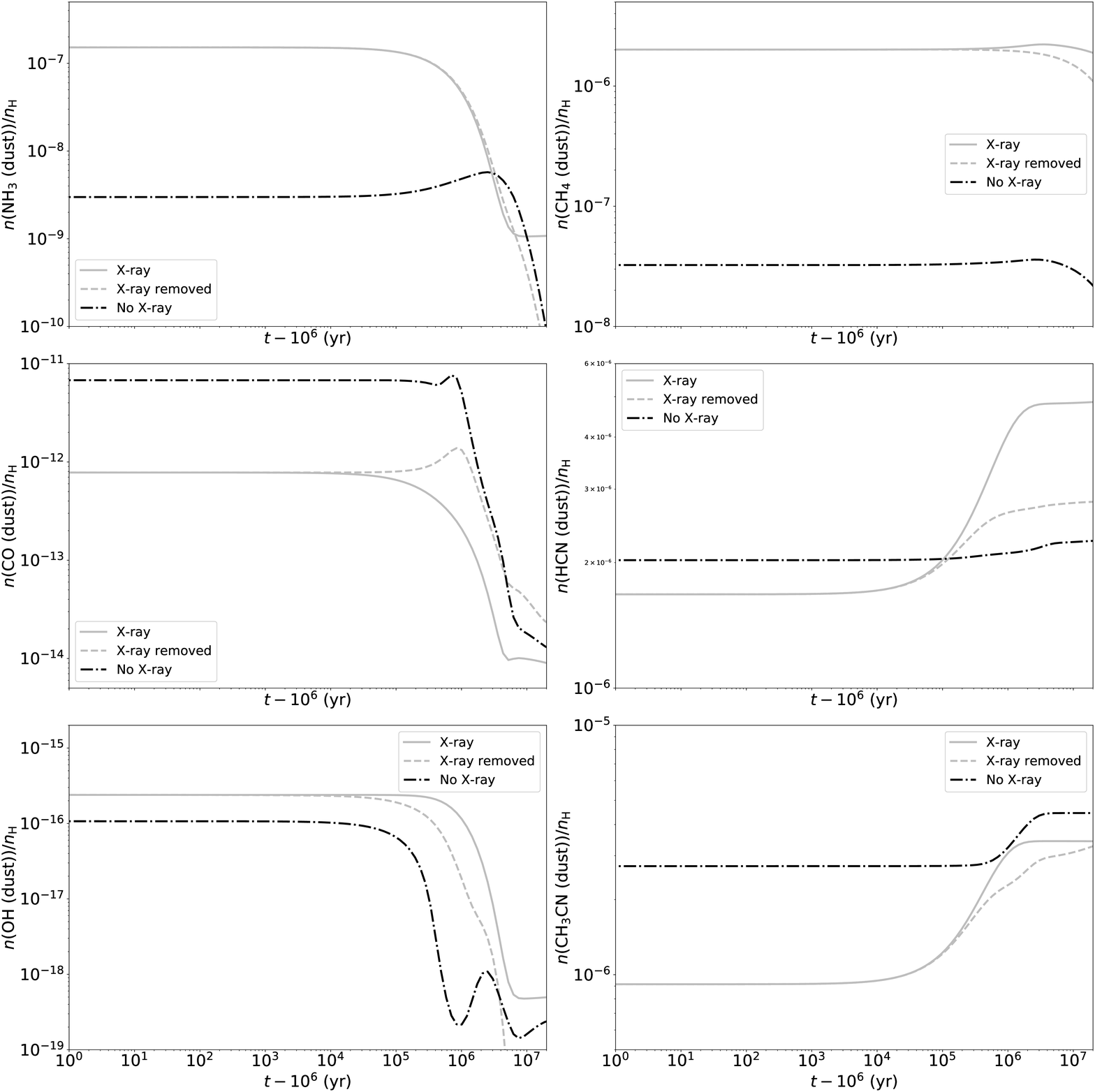}
		\caption{Abundances of \ce{NH3,\ CH4,\ CO,\
		HCN,\ OH,\ CH3CN} on the grain surface as a function of time
		in our fiducial model.}
	\label{fig:app_dust} \end{figure*}

\begin{figure*} \centering
		\includegraphics[width=0.9\linewidth]{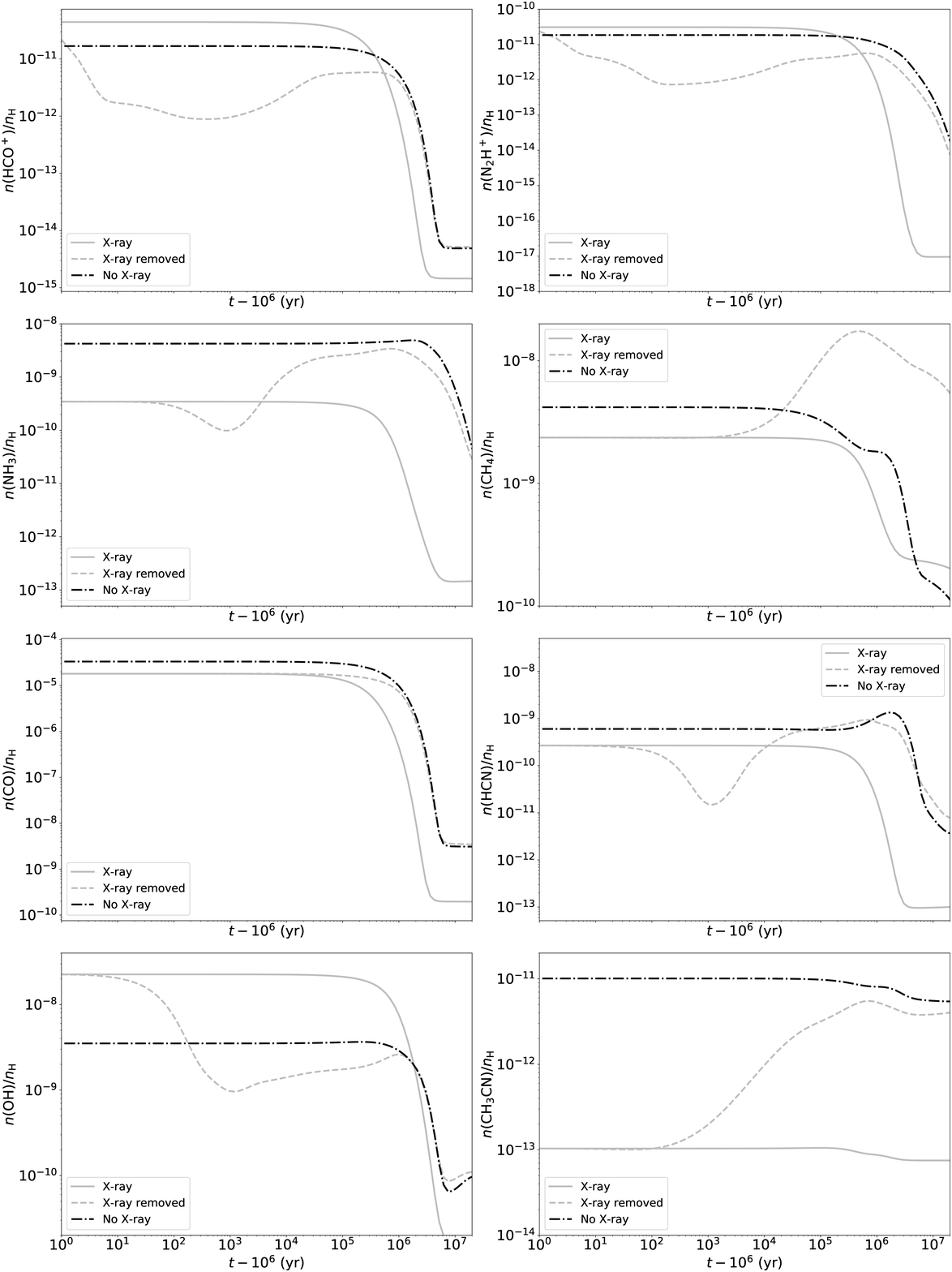}
		\caption{Same as Fig.~\ref{fig:app_dust} but for the molecules
		in the gas phase. Two molecular ions (\ce{HCO+} and \ce{N2H+}) are also included}
	\label{fig:app_gas} \end{figure*}

We also calculated the evolution of several other species that are often
observed in molecular clouds, including two molecular ions (\ce{HCO+,\ N2H+}) and six
neutral species (\ce{NH3,\ CH4,\ CO,\ HCN,\ OH,\ CH3CN}).  The abundances on
the grain surface and in the gas phase are shown, respectively, in
Figures~\ref{fig:app_dust} and \ref{fig:app_gas}. The Calculation is conducted assuming our fiducial condition.

On the grain surface, the abundances of \ce{NH3,\ CH4,\ HCN,\ OH} are
enhanced by X-rays, while those of \ce{CO} and \ce{CH3CN} slightly decrease
in the presence of X-ray irradiation. In the gas phase, most of these molecular
species tend to show lower concentration under X-ray irradiation. When we turn
off the AGN, for most species the gas abundances quickly recover, but
\ce{CH4} shows a significant excess compared to the abundance in the case
without X-rays. \ce{HCO+} and \ce{N2H+} are temporarily produced in higher
efficiency in the first $10^6$ years of X-ray irradiation, and then the
abundances decrease to values lower than those in the case without X-rays.

\begin{figure*} 
\centering
	\includegraphics[width=\linewidth]{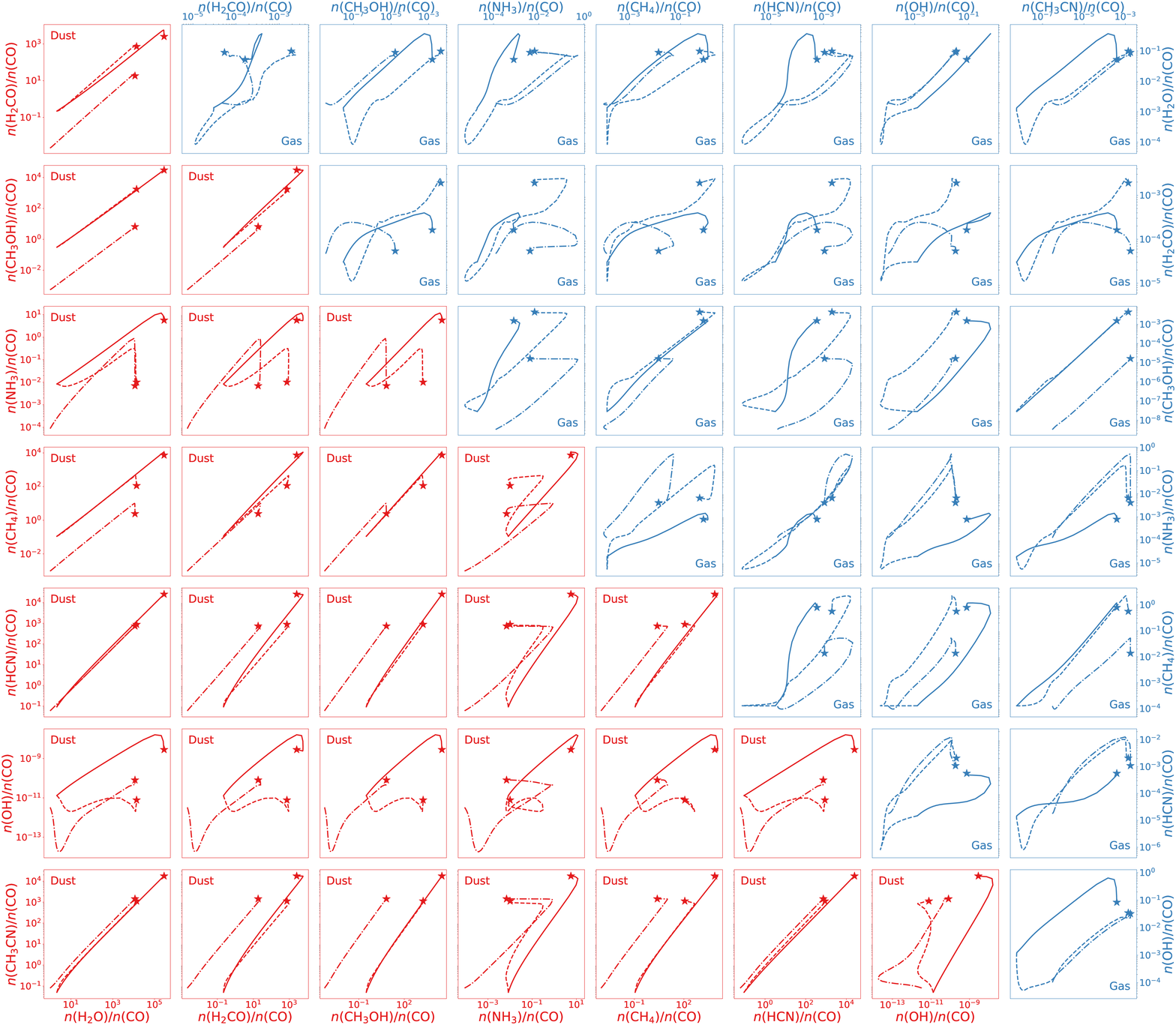} \caption{Same as Fig.~\ref{fig:diag}, but shows all the neutral species listed in \ref{sec:appendix}.
} \label{fig:app_diag} \end{figure*}

For completeness, we also plot in Figure~\ref{fig:app_diag} the
diagnostics using all the neural species shown above. We find that in the gas
phase (blue panels), the evolutionary tracks with and without X-ray irradiation are well separated.
Therefore, we conclude that the gas abundances of the neutral molecules studied in this work
could be good tracers of the past AGN in the Galactic Center. The evolutionary tracks
of the dust species (red panels) are also separated, except for the pairs
\ce{H2CO-CH4}, \ce{H2O-HCN}, \ce{H2O-CH3CN}, and \ce{HCN-CH3CN}. 

\bibliographystyle{apj.bst}

\end{document}